\documentclass[aps,preprintnumbers,prl,twocolumn,showpacs,superscriptaddress]{revtex4}
\usepackage{amssymb}
\usepackage{amsmath}
\usepackage{amsfonts}
\usepackage{amsopn}
\usepackage{graphics}
\usepackage{graphicx}
\usepackage{subfigure}

\begin{document}

\title{Stochastic treatment of finite-$N$ effects in mean-field systems and its application to the lifetimes of coherent structures}

\author{W. Ettoumi}
\affiliation{Ecole Normale Sup\'{e}rieure de Cachan, 94235 Cachan,
France} \affiliation{Laboratoire de Physique des Plasmas CNRS-Ecole
Polytechnique, 91128 Palaiseau cedex, France}
\author{M.-C. Firpo}
\affiliation{Laboratoire de Physique des Plasmas CNRS-Ecole
Polytechnique, 91128 Palaiseau cedex, France}

\date{\today}

\newlength{\textlarg}
\newcommand{\strike}[1]{%
   \settowidth{\textlarg}{#1}
   #1\hspace{-\textlarg}\rule[0.5ex]{\textlarg}{0.5pt}}

   \newcommand{\f}[2]{{\ensuremath{\mathchoice%
    {\dfrac{#1}{#2}}
        {\dfrac{#1}{#2}}
    {\frac{#1}{#2}}
    {\frac{#1}{#2}}
    }}}
\newcommand{\Sum}[2]{\ensuremath{\sum\limits}_{#1}^{#2}}

\begin{abstract}
A stochastic treatment yielding to the derivation of a general
Fokker-Planck equation is presented to model the slow convergence
towards equilibrium of mean-field systems due to finite-$N$ effects.
The thermalization process involves notably the disintegration of
coherent structures that may sustain out-of-equilibrium
quasistationary states. The time evolution of the fraction of
particles remaining close to a mean-field potential trough is
analytically computed. This indicator enables to estimate the
lifetime of coherent structures and thermalization timescale in
mean-field systems.
\end{abstract}


\maketitle


Many physical systems may be considered as isolated assemblies of
$N$ bodies interacting via long-range pair interactions. This is the
case for systems ranging from charged particles interacting via
Coulomb interaction to self-gravitating massive objects like
globular clusters or stars in galaxies and this may even include
suitably prepared Bose-Einstein condensates \cite{ODell} in a close
future. The physically relevant issue of the dynamics of those
systems in the large-$N$ limit forms the subject of kinetic theory.
Long-range systems are prone to collective behavior
that may be largely dominating before binary collisional
effects set on. This hierarchy between collective and collisional behavior
is responsible for the unusual properties of the relaxation process
towards equilibrium as well as for the richness and complexity of
the physics of long-range systems. These are motivations for the
present considerable interest raised by long-range systems in
various fields such as plasma physics, astrophysics
\cite{Gabrielli2010}, statistical physics \cite{Campa} or applied
mathematics.

Collective behavior of long-range systems as well as the
intricacies of the relationships between their dynamics, kinetic
theory and equilibrium statistical properties may be more
conveniently unveiled through models that are already of mean-field
type for finite $N$. These are Hamiltonian models describing e.g.
wave-particle interaction \cite{ElskensEscande}, which is an
ubiquitous phenomenon in hot and dilute plasmas, or the all-to-all
coupling of $N$ bodies in long-range interactions of the kind
of gravitation in a compact space \cite{AntoniRuffo,EttFir}.
Despite their relative simplicity, such models develop a rich long-range
phenomenology.
This includes in particular the emergence of quasistationary states
(QSSs) having lifetimes diverging with $N$, during which the time
average of macroscopic quantities, such as the temperature or the
modulus of mean-fields, differs from their equilibrium statistical
mechanics ensemble averages. These QSSs may be connected to the
existence of coherent structures that may be viewed as long-lived
phase space patterns related to locally insufficient mixing
properties \cite{CoherentFH}. Consequently, the relaxation to
equilibrium should accompany the disintegration of coherent
structures.


The Letter is organized as follows: first, a stochastic treatment of
finite $N$-effects in mean field systems will be proposed leading to
the establishment of a Fokker-Planck equation. In order to test this
model, we then shall consider an Hamiltonian model of $N$ particles
in self-consistent interaction via a cosine potential. Starting from
configurations where $\mathcal{O}(N)$ particles are trapped into
their self-potential well, an analytic expression giving the
fraction of the particles that remain trapped as a function of time
will be successfully tested against numerical results. The relevance
of this indicator to the thermalization issue will be shortly
discussed.

Consider $N$ particles evolving in
the phase space $S_{L}^N \times \mathbb{R}^N$ under the dynamics
deriving from the Hamiltonian
\begin{equation}
\mathcal{H} = \Sum{i=1}{N} \f{{p_i}^2}{2} + \f{1}{2 N} \Sum{n=1}{s}
\Sum{i,j=1}{N} V_n \cos\left[k_n(q_j-q_i)\right], \label{eqn:hamil}
\end{equation}
where $q_i\in S_{L}$ is the position of particle $i$ on the circle
$S_{L}\equiv\mathbb{R}/L$, $p_i$ its conjugate momentum, and where
only the first $s$ long-range components with wave numbers $k_n=2\pi
n/L$, for $1\leq n \leq s$, are retained in the potential term. When
$V_n \propto n^{-2}$, model (\ref{eqn:hamil}) amounts to the
long-range truncation of the one-dimensional Newtonian potential
with space-periodic boundary conditions, that describes Coulomb or
gravitational interaction depending on the potential sign. Various
systems covered by (\ref{eqn:hamil}) have been discussed in
Ref.~\cite{Elskens}. An extension to a spatial dimension $d>1$
should not be a conceptual problem. Introducing the set of
collective observables $\{\mathbf{M}_n\}$ through
\begin{equation}
\mathbf{M}_n= \f{1}{N}\Sum{j=1}{N}\left( \cos(k_n q_j), \sin(k_n
q_j)\right)=M_n (\cos \phi_n, \sin \phi_n) \label{eqn:def_mn}
\end{equation}
yields the equation of motion of any particle $i$ as
\begin{equation}
\ddot{q}_i + \Sum{n=1}{s} k_n V_n M_n \sin(k_n
q_i - \phi_n) = 0. \label{general_pendulum}
\end{equation}
Therefore, the collective variables $M_{n}$ behave as mean fields
that, as well as the phases $\phi_n$, depend on time through the
self-consistency relations (\ref{eqn:def_mn}).

For smooth potentials like in (\ref{eqn:hamil}), the convergence of
the finite-$N$ dynamics to Vlasov equation is rigorously proved on
arbitrary finite-time intervals \cite{Spohn}. Vlasov equation being
time reversible, its solution $f(p,q,t)$ cannot approach an
equilibrium \cite{VanKampen}, yet macroscopic quantities, involving
phase space integrals of $f$, such as the mean-fields, can converge
to stationary values associated to QSSs. In the realistic finite-$N$
Hamiltonian framework, finite-$N$ effects will in the long term
induce the thermalization process and disintegration of coherent
structures possibly sustaining those QSSs.

\begin{figure}[htbp]
\begin{center}
     \subfigure{\includegraphics[scale=0.28]{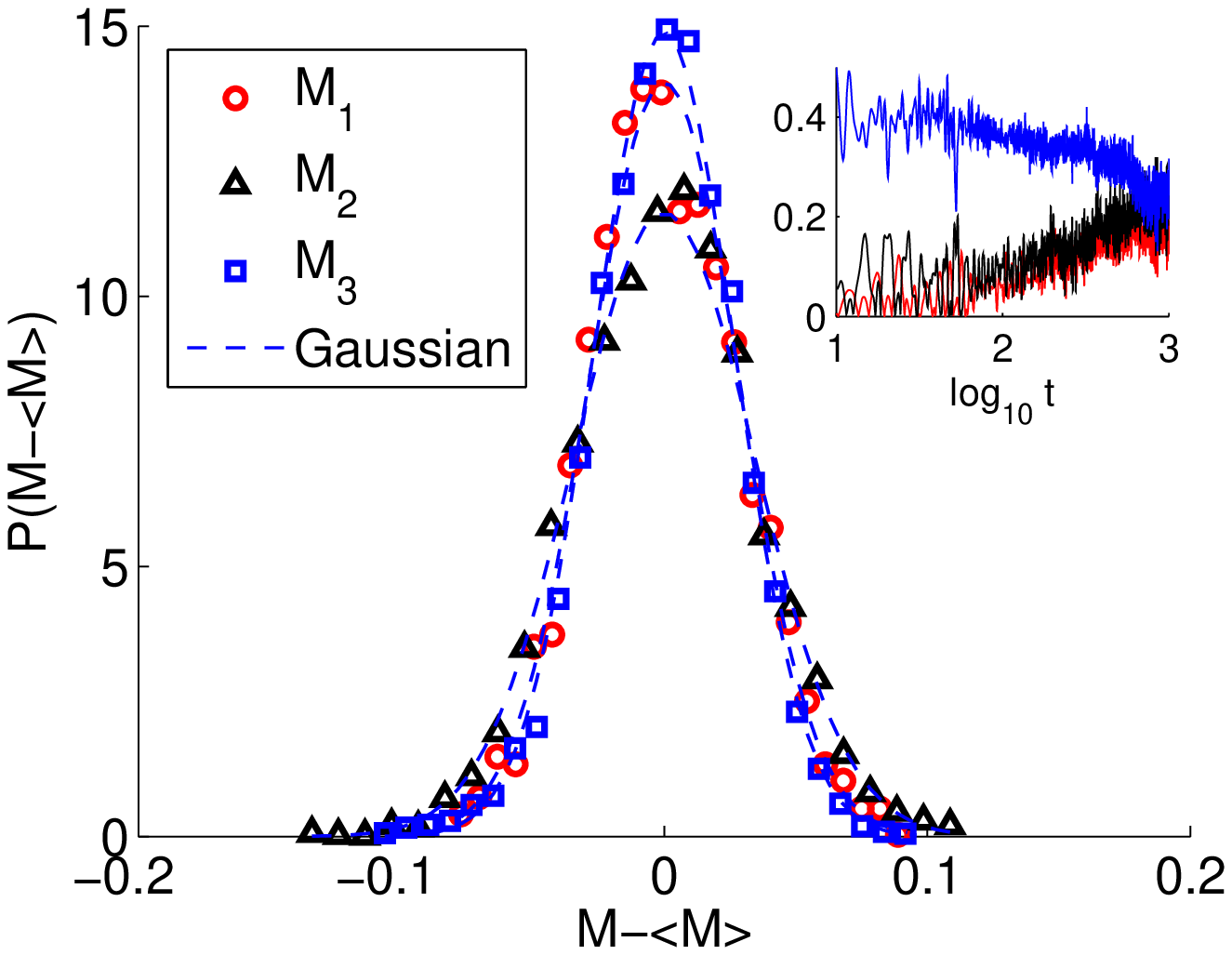}}
  \subfigure{\includegraphics[scale=0.28]{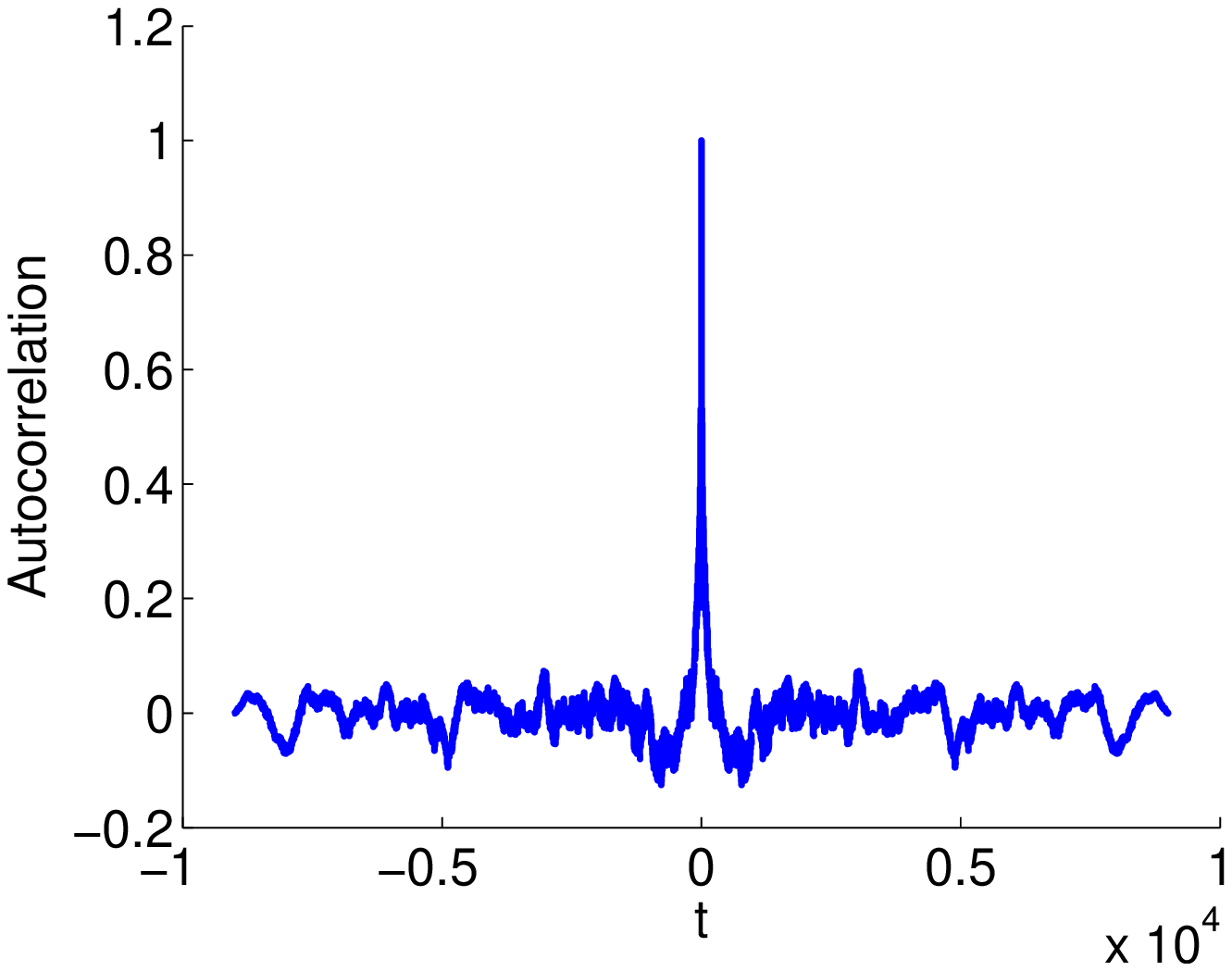}} \\
\end{center}
\caption{(Left) Comparison between the numerically computed
fluctuations of $M_{1}$, $M_{2}$ and $M_{3}$ around their local
average $\left\langle M_n\right\rangle$ with a Gaussian fit, for
system (\ref{eqn:hamil}) with $s=3$ and $V_n=n$ starting from an
arbitrary initial condition. The inset shows the evolution of $M_1$,
$M_2$ and $M_3$ with respect to time. (Right) Autocorrelation of
$M_1$ after subtracting its local average.}
\label{fig:fluctuations_hist}
\end{figure}

Modelling this process, we assume that the system
(\ref{eqn:hamil}) is trapped in a QSS, such that one can write the
mean fields as $M_{n}(t) = M_{n}^0(t) + \delta M_{n}(t)$ with
$\delta M_{n} \ll M^0_{n}$, where $\delta M_{n}$ varies on a time
scale which is very much smaller than the characteristic time scale
of $M_{n}^0$, the latter being comparable to a local average. We now
make the hypothesis that during the QSS regime, the fluctuations
around the local mean value have a variance decreasing as $N^{-1}$.
The central idea behind this is to replace the deterministic but yet
very chaotic fluctuations of $\delta M_n$ by stochastic processes,
whose variances are suitably chosen. Numerical observations support
this modeling. For instance, as shown in
Fig.~\ref{fig:fluctuations_hist} for some special case, the mean
fields clearly exhibit two different timescales, in agreement with
the decomposition suggested earlier. Moreover, both histograms
and the autocorrelation function shown in
Fig.~\ref{fig:fluctuations_hist} suggest that one can model the
fluctuations $\delta M_n$ by a Gaussian (white) noise.

Replacing $\delta M_{n}$ by the noise $\xi_n$ such that
$\left\langle \xi_n \right\rangle$=0, and
$\left\langle\xi_n(t)\xi_n(t')\right\rangle\propto N^{-1}
\delta(t-t')$, the Fokker-Planck equation (FPE) associated to the
Langevin equations coming from the stochastic version of the
equations of motion~(\ref{general_pendulum}) reads
\begin{multline}
\f{\partial f}{\partial t} + \f{\partial}{\partial q}\left(p f\right)
-\Sum{n=1}{s} k_n V_n M^{0}_{n} \sin(k_n q - \phi_n)\f{\partial f}{\partial p} \\
=\Sum{n=1}{s} \f{{k_n}^2{V_n}^2}{2}
\left\langle{\xi_n}^2\right\rangle \sin^2(k_n q-\phi_n)
\f{\partial^2 f}{\partial p^2} \label{eqn:FPE}
\end{multline}
This equation may be interpreted as a Vlasov equation supplemented
with a r.h.s. of order $N^{-1}$, consistently with the argument
presented in Ref.~\cite{Campa}, coming here not from binary
collisions but from the fluctuations of the mean fields. This
differs from other FPEs derived in mean-field systems in other
places \cite{BouchetDauxois2005,Chavanis2010}. Moreover, this
equation was derived without any need to invoke dissipation (see
e.g. \cite{Mallick}).

\begin{figure}[htbp]
\centering \subfigure{\label{fig:M_N1_N2-a}
\includegraphics[width=0.3\columnwidth,height=22mm]{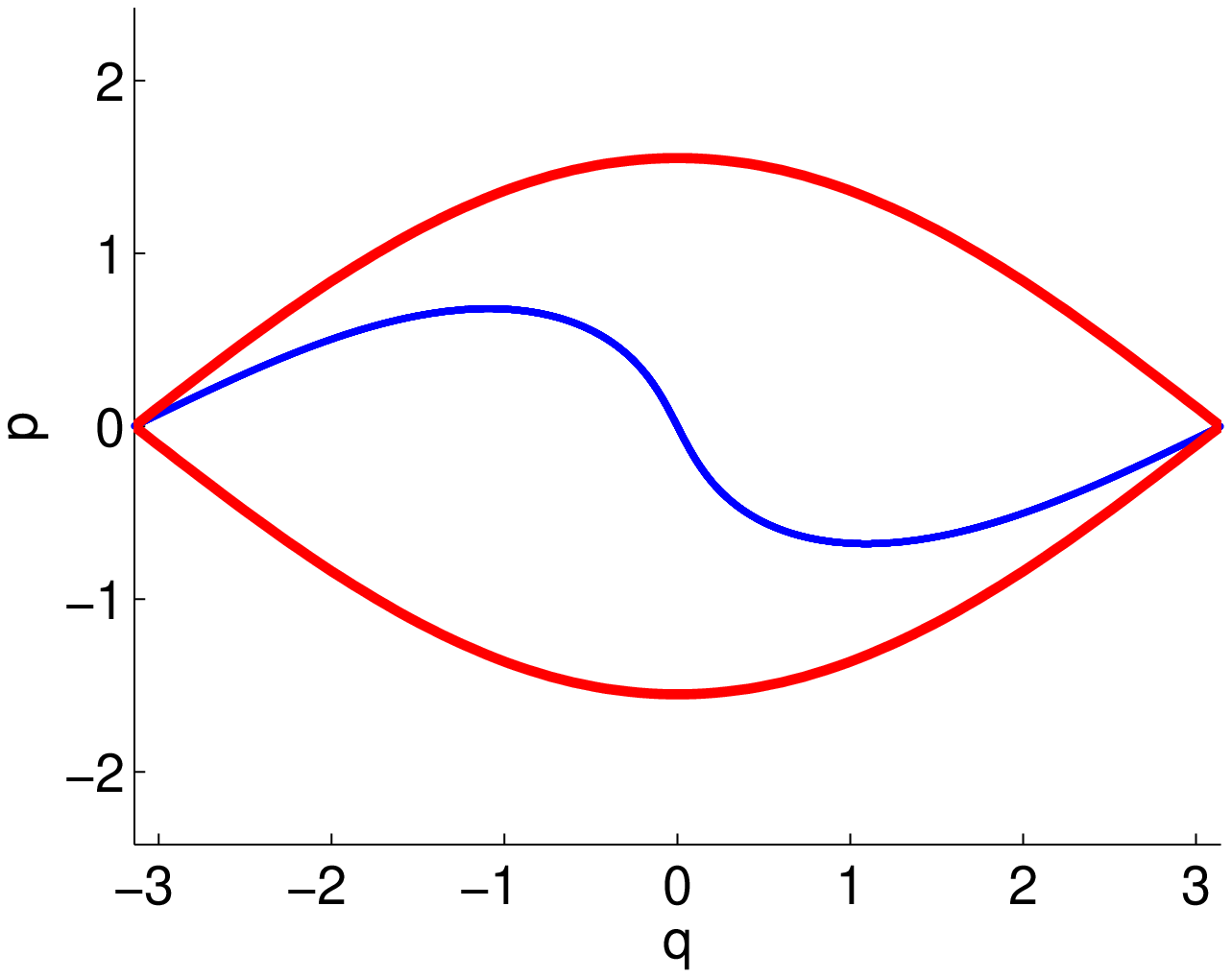}}
\subfigure{\label{fig:M_N1_N2-b}
\includegraphics[width=0.3\columnwidth,height=22mm]{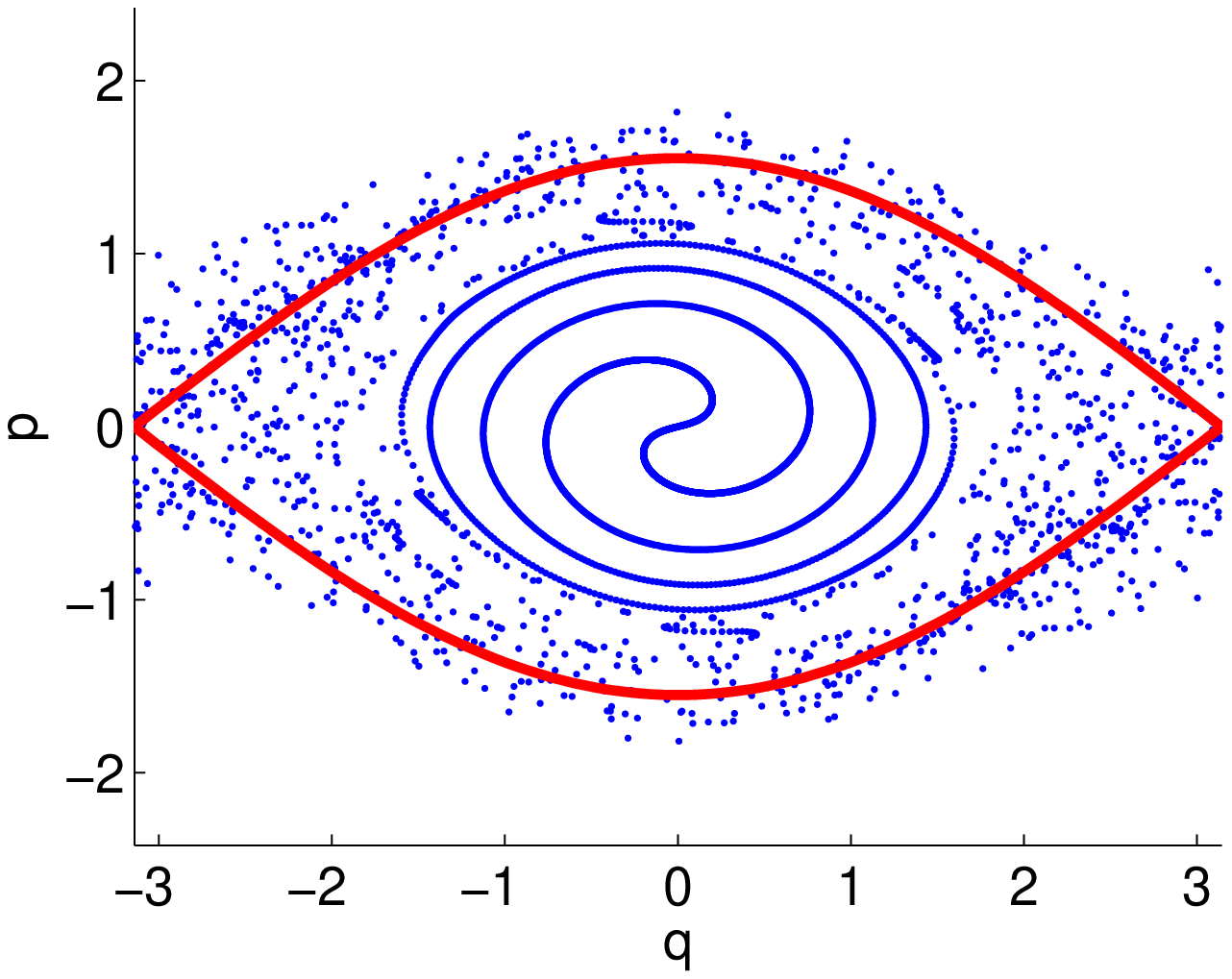} }
\subfigure{\label{fig:M_N1_N2-c} \includegraphics[width=0.3\columnwidth,height=22mm]{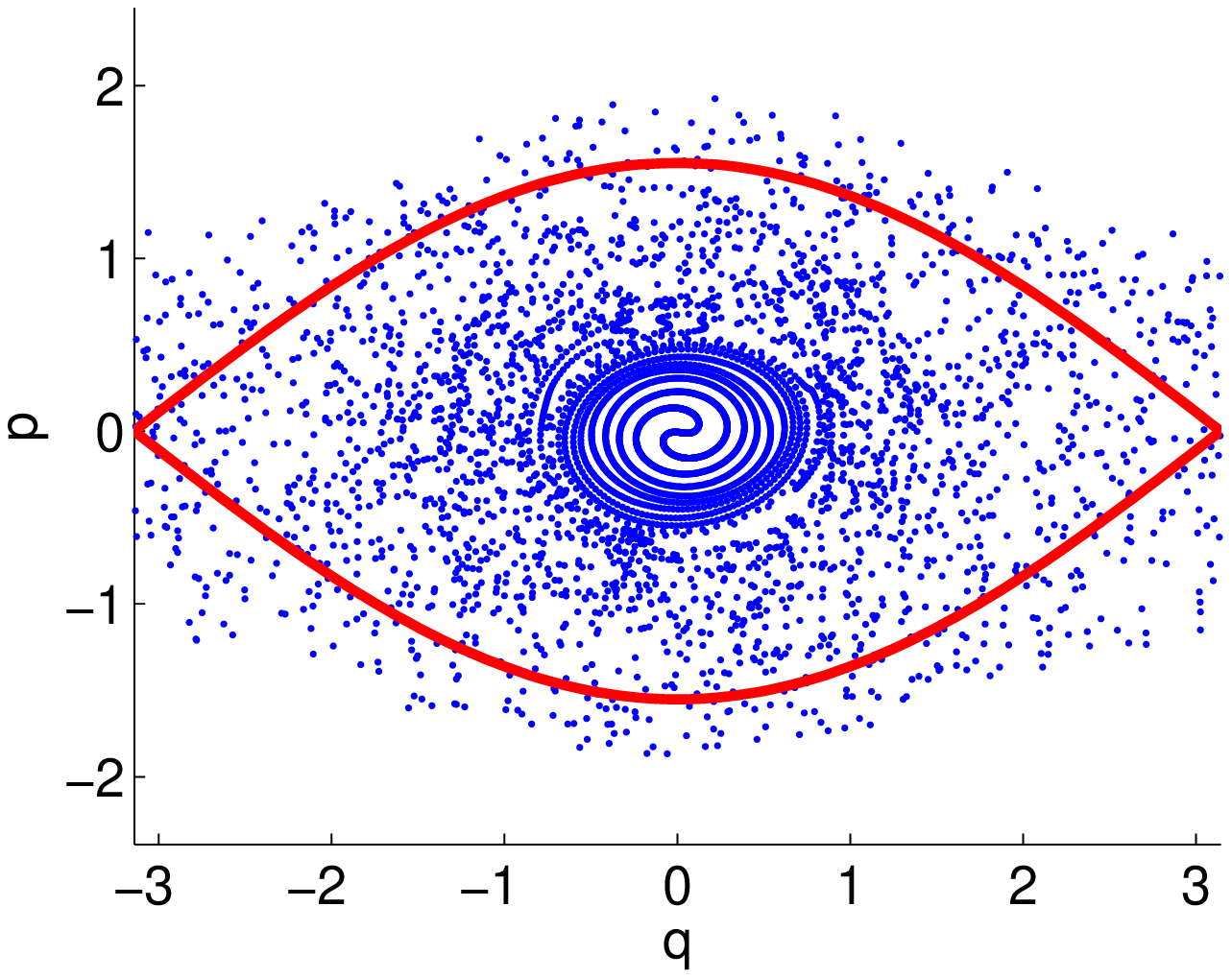} }\\
\subfigure{\label{fig:M_N1_N2-d}
\includegraphics[width=0.98\columnwidth,height=45mm]{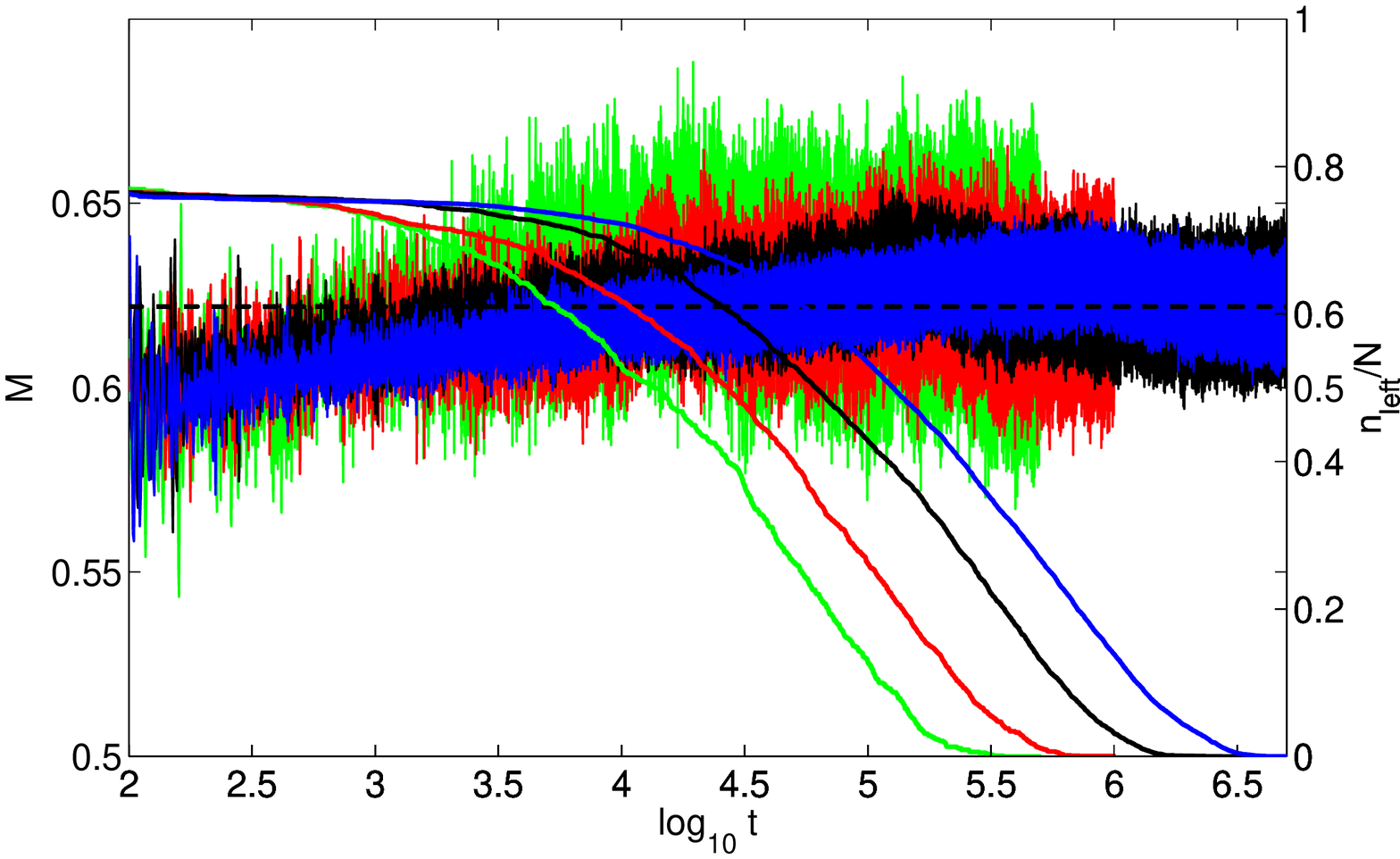}}
\caption{(Upper panel) Phase-space snapshots at increasing times (from left to right, $t=6$, $t=100$ and $t=2000$) for an initial cold-beam condition of energy $U=0.5$. The simulation was performed using $N=10^4$ particles.The instantaneous separatrix is plotted in bold red. (Lower panel) Time
evolution of the magnetization $M$ and fraction $n_\ell/N$ for
$N=10^3$ (green), $2.10^3$ (red), $5.10^3$ black and $10^4$ (blue)
particles.} \label{fig:M_N1_N2}
\end{figure}

In what follows, we shall consider the case of a single resonance
($s=1$) in which coherent structures may survive for long times
close to the potential trough (see e.g. Fig.~\ref{fig:M_N1_N2}). The
FPE (\ref{eqn:FPE}) may be further simplified by looking for
solutions in separate variables $q$ and $p$ writing
$f(p,q,t)=g(q,t)\tilde{f}(p,t)$. Assuming that $g$ is even in the
wave frame and that $\int_{0}^{L} g(q,t)\mathrm{d}q$ is constant,
one obtains a simple diffusion equation
\begin{equation}
\f{\partial \tilde{f}}{\partial t} = \mathcal{D}(t) \f{\partial^2
\tilde{f}}{\partial p^2}, \label{eqn:FPE_avg_sym}
\end{equation}
with diffusion coefficient
\begin{equation}
\mathcal{D}(t)= \f{{k}^2 {V}^2}{2} \left\langle{\xi}^2\right\rangle
\overline{\sin^{2}(k q - \phi)}, \label{eqn:FPE_avg_gnonzero}
\end{equation}
and $\overline{\sin^{2}(k q - \phi)}\equiv \int_{0}^{L} \sin^{2}(k q
- \phi)g(q,t)\mathrm{d}q / \int_{0}^{L} g(q,t)\mathrm{d}q$.
Numerical evidence \cite{Campa} supports the fact that $p$ and $q$
may be treated as separate variables and that $q$ may be considered
as a fast variable compared to $p$, meaning that the distribution
function in $q$ approaches much more quickly its Boltzmann-Gibbs
shape than the $p$ one consistently with basic dimensional arguments
\cite{Mallick}. The forthcoming numerical tests will show that the
average of $\sin^{2}(k q - \phi)$ may be effectively replaced by its
ensemble average.

From now on, in order to simplify expressions, we shall put $V=1$
and $L=2\pi$, which gives $k=1$. This amounts to the well-known
Hamiltonian Mean Field model \cite{Campa}. Figure \ref{fig:M_N1_N2}
shows numerical results obtained starting from a monokinetic beam
\cite{EttFir}. The upper panel shows the one-particle phase space
plots at three different stages of the evolution. Deep inside the
mean-field potential trough, particles move almost regularly forming
a clear coherent structure pattern that progressively dissipates. It
is interesting to note the similarity of these figures with the
phase space plots for the 1-D finite-$N$ cold dark matter
simulations of Ref. \cite{Binney2004}. The lower panel shows the
evolution of the mean-field for four different numbers of particles,
ranging from $N=10^{3}$ to $N=10^{4}$. The dashed line marks its
equilibrium ensemble average. It is clear from the figure that the
convergence towards equilibrium slows down as $N$ increases. Also
displayed is the evolution of $n_\ell/N$, that is the fraction of
particles initially trapped within the separatrices and which remain
inside them up to time $t$. During a short transient regime (not
shown), about twenty percent of the particles escape the mean-field
resonance. It is only after this lapse of time that the regime
becomes diffusive. Most importantly, Fig.~\ref{fig:M_N1_N2} gives an
evidence that the cancelation time of $n_\ell/N$ may be used as a
marker of thermalization as it qualitatively coincides with the time
where the mean-field begins to fluctuate around its equilibrium
value in a stable way. This is not surprising since, at the time
when $n_\ell/N$ vanishes, the coherent structure has been completely
disintegrated. The dependence of $M$ on $N$ becomes much clearer when
increasing $N$ to values of the order of $10^5$, but the numerical
cost was too important for us to perform a full simulation leading to
a vanishing $n_\ell/N$. However, the early-time behaviour of $n_\ell/N$
was still correctly described by our model.

The equations of the separatrices are given by $p_{S}(t)=\pm
2\sqrt{M(t)}\cos (q/2)$. Let us put $\lambda = 2  \sqrt{M^0}$ and
consider it to be almost constant. Figure \ref{fig:M_N1_N2} is a
motivation to answer the following question: What is the number of
particles $n_\ell(t)$, having initially momenta comprised between
$-\lambda$ and $\lambda$, that remain in this domain up to time $t$
? Among such particles are the particles forming the core of
coherent structures that slow down the mixing process. However,
because of the parametric resonance induced by the mean fields
fluctuations, these particles will eventually escape, and we shall
now estimate the characteristic time needed by the system to
evacuate a fraction $1-\delta$ of the $N_0$ particles initially
contained in the band of momenta $[-\lambda;\lambda]$. Using the
linearity of the diffusion equation (\ref{eqn:FPE_avg_sym}), one can
focus on the contribution of particles whose momenta remain up to
time $t$ within the band $[-\lambda;\lambda]$ and solve
(\ref{eqn:FPE_avg_sym}) by imposing the cancelation of $f$ at $p=\pm
\lambda$. Finally, integrating over $p$ yields the solution
\begin{equation}
\frac{n_\ell(t)}{N_{0}} = \Sum{n=0}{\infty} \f{2 (-1)^{n+1}
c_{n}}{(2n+1) \pi} \mathrm{exp}\left(-\f{(2n+1)^2\pi^2}{4 \lambda^2}
\mathcal{D} t\right), \label{eqn:nl}
\end{equation}
with $\mathcal{D}$ given by Eq. (\ref{eqn:FPE_avg_gnonzero}) and
\begin{equation}
c_{n}=\frac{\int_{-\lambda}^{\lambda}\tilde{f}(p,t=0)\cos\left(\frac{(2n+1)p\pi}{2
\lambda}\right)\mathrm{d}p}{\int_{-\lambda}^{\lambda}\tilde{f}(p,t=0)\mathrm{d}p}.
\label{coef:ck}
\end{equation}

\begin{figure}[htbp] \centering
{\includegraphics[width=0.35\columnwidth,height=30mm]{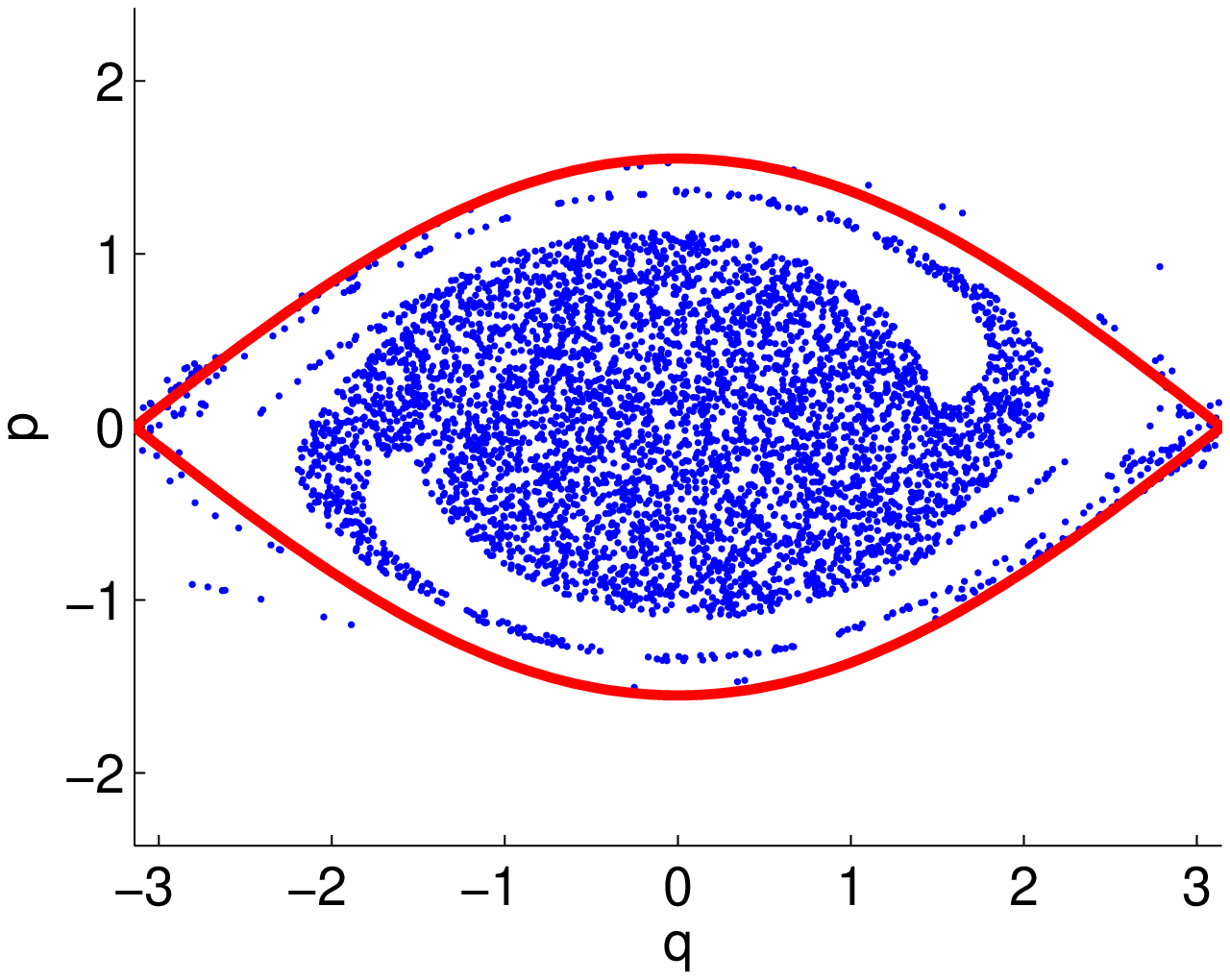}
\label{fig:WB-a}}
{\includegraphics[width=0.62\columnwidth,height=30mm]{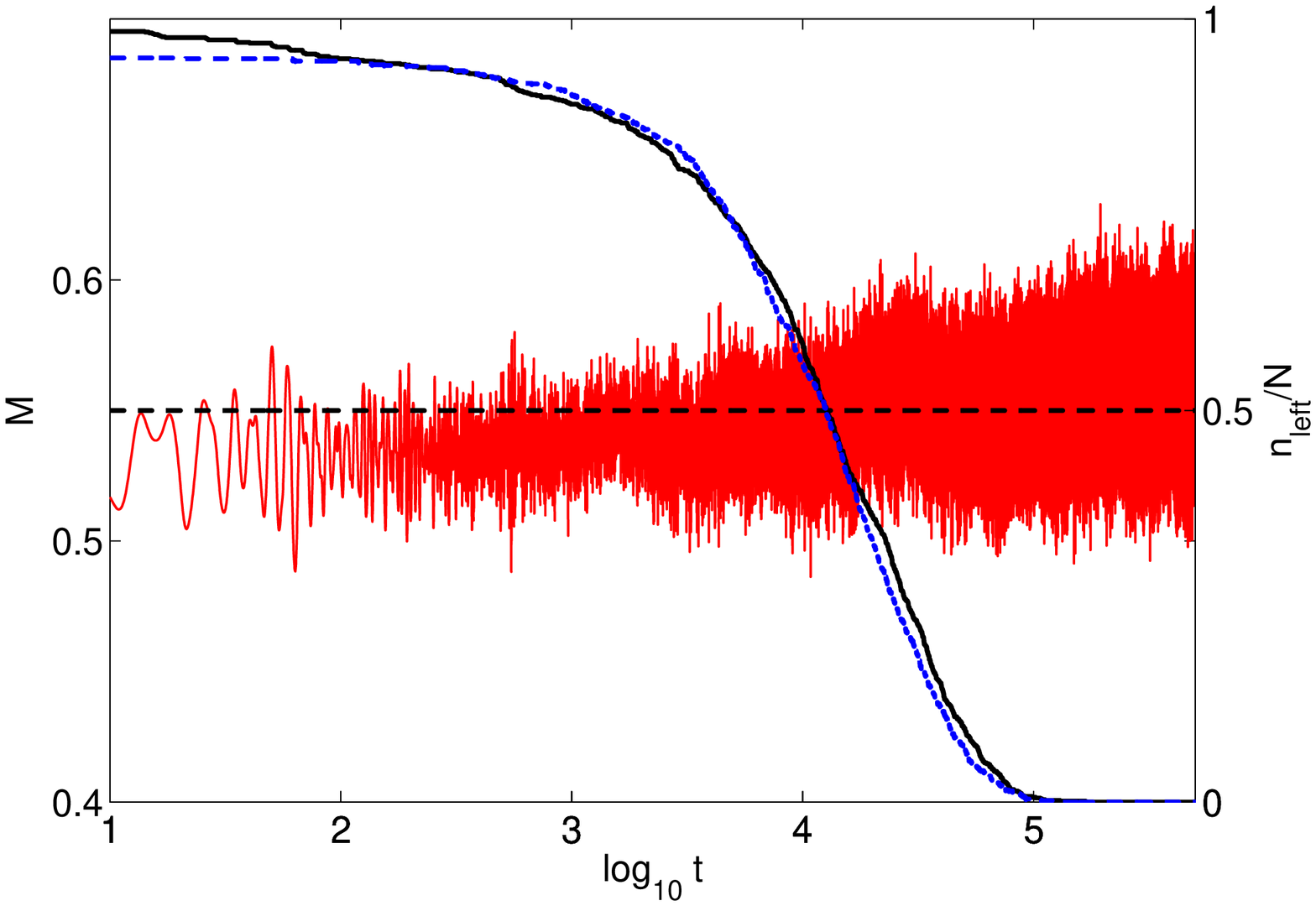} \label{fig:WB-b}}\\
\caption{(Left) Snapshot of the one-particle phase space at $t=12$
for $N=5000$ particles initially distributed in a waterbag
configuration (\ref{WB}) with $\Delta p=0.848$ and $\Delta q =
2.16$. (Right) Time evolution of the mean-field $M$ (red curve) and
of $n_{l}/N$ (black curve). The blue dashed curve is the analytic
expression for $n_{\ell}/N$ as deduced from Eqs.
(\ref{eqn:nl})-(\ref{coef:ck}). The initial time has been chosen at
time $t=10$ and $N_{0}=0.95 N$.} \label{fig:WB}
\end{figure}

As confirmed by the numerical simulations shown in
Fig.~\ref{fig:WB}, the dynamics arising from Eq.
(\ref{eqn:FPE_avg_sym}) proves to correctly depict the escape
process through the validation of Eqs.
(\ref{eqn:nl})-(\ref{coef:ck}). In these numerical simulations,
particles were initially distributed according to a so-called
waterbag distribution
\begin{equation}
f_{0}(p,q)=\Theta(\Delta p -|p|)\Theta(\Delta q -|q|)/(4 \Delta p
\Delta q) \label{WB}
\end{equation}
where $\Theta$ stands for the Heaviside step function. It is
interesting to note that when starting from these initial waterbag
conditions, the phase-space distribution eventually exhibits a
core-halo structure, which has been recently investigated in
\cite{Pakter2011}.  Once again, it is visible on Fig.~\ref{fig:WB}
that, at the time when all the particles that where initially in the
momentum band $[-\lambda;\lambda]$ have at least once escaped this
domain, the system has seemingly reached its thermal equilibrium. In
order to test the escape model given by Eqs.
(\ref{eqn:nl})-(\ref{coef:ck}), one needs to know the diffusion
coefficient $\mathcal{D}$ given by Eq. (\ref{eqn:FPE_avg_gnonzero}).
As already discussed, $\overline{\sin^{2}(q)}$ may be estimated from
$\left\langle \sin^{2} q \right\rangle_c = I_1\left(\beta
\left\langle M\right\rangle_c\right)/[\beta \left\langle
M\right\rangle_c I_0\left(\beta \left\langle
M\right\rangle_c\right)]$. A priori $\langle \xi^{2}\rangle$ has to
be determined from numerical simulations since the system is not at
equilibrium. However, in the cases that were considered, the
numerically computed variance $\langle \xi^{2}\rangle$ was almost
indistinguishable from its canonical value given by
\begin{equation}
\left\langle \delta^2 M \right\rangle_c = \f{2}{N}
\f{\partial}{\partial \beta} \log \left[\f{v^*}{\beta} \sqrt{\f{2\pi
N}{\partial^2_v \psi|_{v^*}}}\mathrm{e}^{-N
\psi(v^*)}\right]-\f{{v^*}^2}{\beta^2}, \label{micro_fluc}
\end{equation}
where $\psi(v)=v^2/(2\beta)-\log I_0(v)$ and $v^*=\beta \left\langle
M \right\rangle_c$ satisfies the self-consistency equation
$\partial_v \psi|_{v^*}=0$. Numerically, this gives $\left\langle
\xi^2 \right\rangle \approx 0.43/N$, which is consistent with the
fit obtained from numerical simulations. As shown in
Fig.~\ref{fig:WB}, the agreement between the numerically computed
time evolution of $n_{\ell}/N$ and its analytic modeling
(\ref{eqn:nl})-(\ref{coef:ck}) is quite satisfactory.

Let us finally estimate the time needed to destroy the inner
coherent structure by the means of Eq.~(\ref{eqn:nl}). Considering
that $N_0=\mathcal{O}(N)$ and $\mathcal{D}=\mathcal{O}(N^{-1})$, a
rough estimate of the time $\tau_\delta(N)$ needed for $n_\ell/N$ to
reach down a sufficiently small fraction $\delta$ gives
\begin{equation}
\tau_\delta(N) \propto - N \log \delta. \label{eqn:tau_delta}
\end{equation}
When $\mathcal{D}$ depends on time, Eq.~(\ref{eqn:tau_delta})
follows from the mean value theorem. This first result recovers the
linear $N$ scaling found in the abundant literature of long-range
interacting systems \cite{AntoniRuffo,Joyce,Chavanis}, where the
numerical evidence is extracted from thresholds equivalent to the
$\delta$ criterion imposed here. One does not expect the continuum
approach behind Eq.~(\ref{eqn:nl}) to remain valid for vanishingly
small values of $\delta$. However, going up to the limit of validity
of this model, one may infer that the sweeping of phase space is
sufficient to reach a complete thermalization at a time $\tau_{QSS}$
when $n_\ell = \mathcal{O}(1)$. Then Eq.~(\ref{eqn:tau_delta}) would
give the maximal scaling $\tau_{QSS} \propto N \log N$. This
corresponds to the scaling recently suggested in Ref. \cite{Gupta}
for the $s=1$ case.
\begin{figure}[htbp]
\begin{center}
  \subfigure{\includegraphics[scale=0.28]{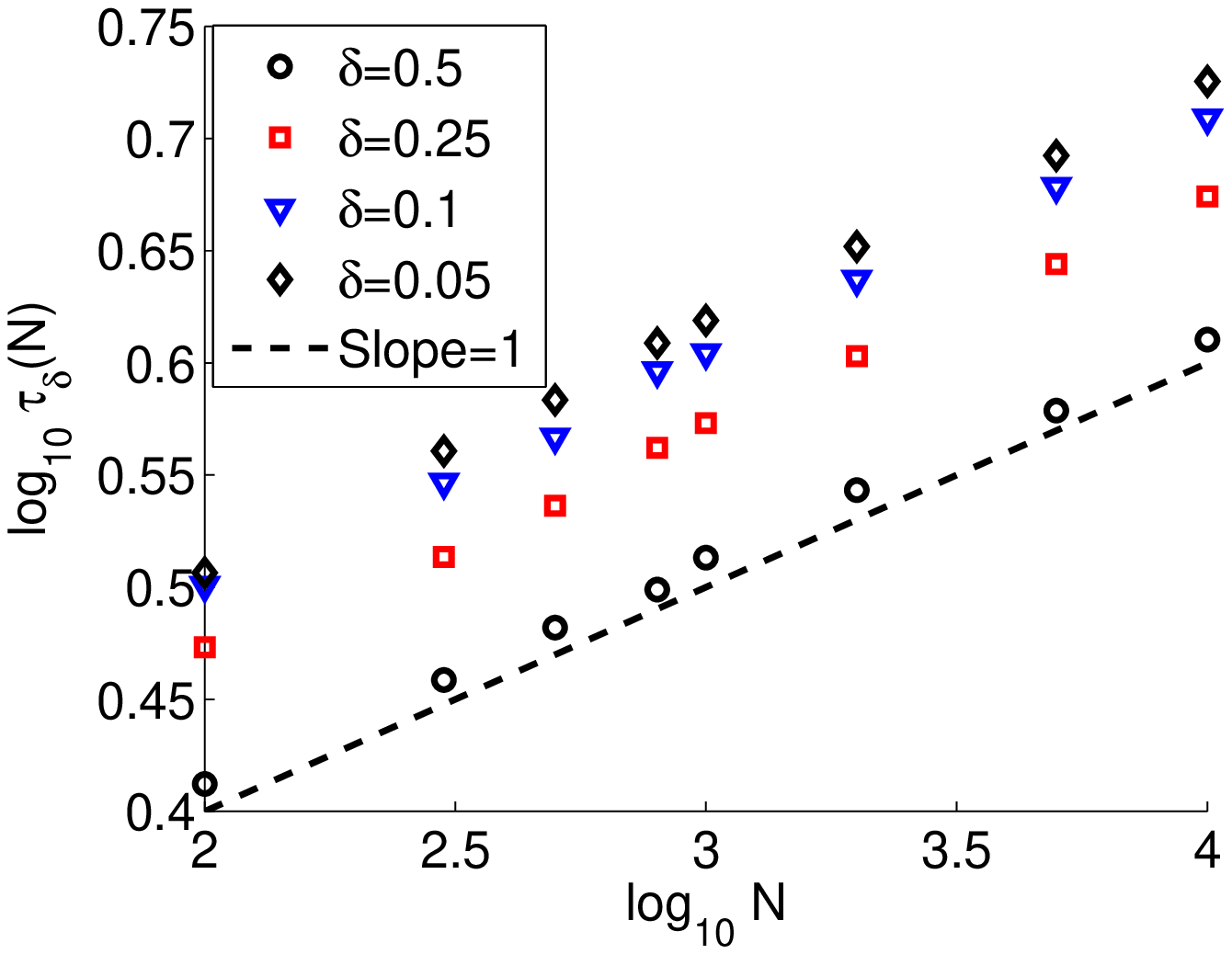}}
  \subfigure{\includegraphics[scale=0.28]{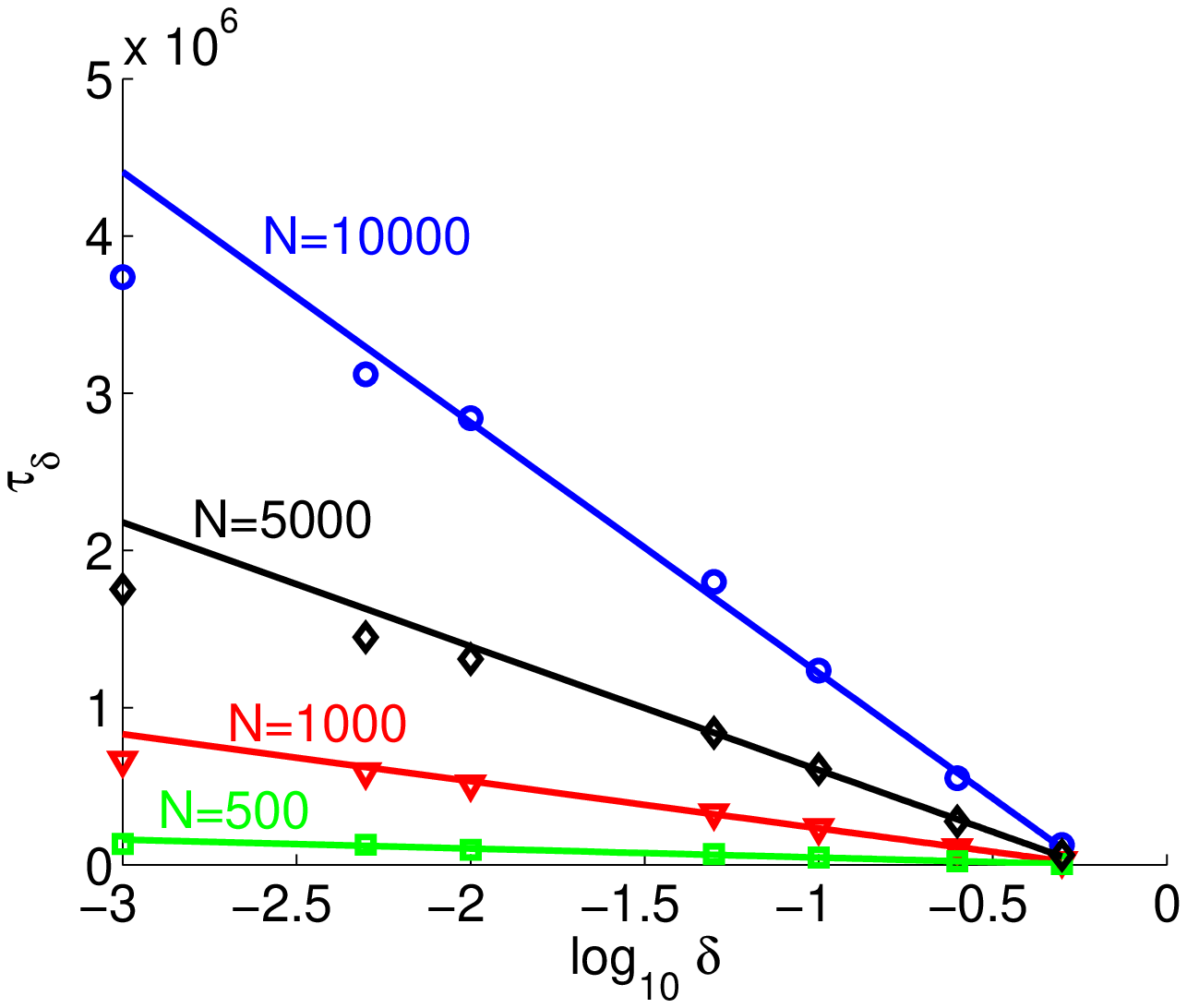}}\\ 
\end{center}
\caption{(Left) Plot of $\tau_\delta(N)$ in log-log scale with
respect to $N$. As predicted by Eq.~(\ref{eqn:tau_delta}), the
behavior is linear with $N$. (Right) Plot of the numerically
measured $\tau_\delta$ with respect to $\log \delta$. The behavior
is linear over a wide range of $\delta$. As expected, the measured
times for very low threshold values are lower than the logarithmic
prediction of Eq.~(\ref{eqn:tau_delta}), since the latter is
obtained in the continuous limit.} \label{fig:scaling_delta}
\end{figure}
Eq.~(\ref{eqn:tau_delta}) predicts a linear behavior with respect to
$N$, which we found to be correct over the whole range of values of
$N$ studied here, independently from the chosen threshold $\delta$.
We also checked the scaling with the latter parameter. Figure
\ref{fig:scaling_delta} shows a very good agreement between
Eq.~(\ref{eqn:tau_delta}) and the numerical simulations.

These scalings contrast with the numerically obtained $N^{1.7}$
scaling for the QSS lifetime starting from two special initial
conditions \cite{Zanette2003,Yamaguchi}. These cases are however not
in contradiction with the results presented here since, even this is
less obvious for \cite{Zanette2003}, they both correspond to QSS
about a vanishing mean-field, a case that is excluded from the
present framework since the phase would be no longer defined.
In the intermediate cases, where the QSS magnetization is clearly
above zero, but yet far from the equilibrium expectation, this method
provides a good estimation of the time needed to destroy the coherent
structures, but fails to predict the QSS lifetime, since the effective
model does not capture the average growth of the separatrix with time.

The present framework and results are expected to be easily
transposable to wave-particle models in which finite-$N$ effects
eventually drive the system towards equilibrium in contradiction
with the Vlasov approach \cite{Firpo01}. This discreteness effect
may be more than a numerical concern for simulations since some
physical effects \cite{YoonPRL2005} cannot be explained in the
Vlasov limit.

\bibliographystyle{unsrt}

\end{document}